\documentstyle[amssymb,twocolumn,aps,prl,epsfig]{revtex}

\twocolumn

\begin{document}
\author{A. Lezama$^{1,2}$, G.C. Cardoso$^1$, and J.W.R. Tabosa$^1$}
\address{$^1$ Departamento de F\'{i }sica, Universidade Federal de Pernambuco,\\
50670-901 Recife, PE - Brazil \\
$^2$ Instituto de F\'{\i }sica, Facultad de Ingenier\'{\i }a, C. Postal.30.\\
11000, Montevideo, Uruguay}
\date{\today }
\title{Polarization dependence of four-wave mixing in a degenerate two-level system.}
\maketitle

\begin{abstract}
Nearly degenerate four-wave mixing (NDFWM) within a closed degenerate
two-level atomic transition is theoretically and experimentally examined.
Using the model presented by A. Lezama et al [Phys. Rev. A 61, 013801
(2000)] the NDFWM spectra corresponding to different pump and probe
polarization cases are calculated and discussed. The calculated spectra are
compared to the observation of NDFWM within the $6S_{1/2}(F=4)\rightarrow
6P_{3/2}(F=5)$ transition of cesium in a phase conjugation experiment using
magneto optically cooled atoms.
\end{abstract}



\section{Introduction.}

The generation of new optical fields is the most spectacular consequence of
the intrinsic non-linearity of the light-matter interaction. Four-wave
mixing (FWM), i.e. the generation of a fourth field as the result of the
interaction of a material sample with three electromagnetic fields is the
simplest non-linear optics generation process allowed in every material\cite
{NICO,SHEN}.

This paper is concerned with nearly degenerate four wave mixing (NDFWM)
occurring when two of the three incident fields (designated as pump fields)
have the same frequency $\omega _1$ and the third field (probe field) has an
independent frequency $\omega _2\equiv \omega _1+\delta $ which is tunable
around $\omega _1$. Of particular interest is the case where all fields are
nearly resonant with the same optical atomic transition. In such case the
spectroscopic study of the NDFWM signal as a function of the probe to pump
frequency offset can provide useful information about the atomic dynamics.
NDFWM was used by Rothberg and Bloembergen\cite{ROTHBERG} for studying
collisional dynamics in sodium vapor. Boyd et al\cite{BOYD} have studied
theoretically the atomic polarization and the propagation of the probe and
NDFWM waves in the case of a pure two-level system (PTLS) driven by a
arbitrarily intense pump. Their work illustrate the critical influence of
the atomic damping mechanism in the FWM process. Steel et al \cite{DUNCAN}
explored NDFWM in both open and closed PTLS. Berman and coworkers \cite
{BERMAN-STEEL} clearly established the suitability of NDFWM for the
determination of all relevant relaxation parameters. Andersen and coworkers
analyzed the NDFWM process in PTLS in terms of the dressed state picture\cite
{ANDERSEN}. Their work stresses the importance of the initial atom-field
state on the FWM yield.

In spite of the fact that the experiments on NDFWM generally involve
degenerate energy levels, most experiments to date were carried under
conditions where the level degeneracy does not play an essential role. In
such cases, simplified theoretical models using two-level schemes were
successfully used. Nevertheless, the role of the level degeneracy for NDFWM
was clearly appreciated by Berman and coworkers \cite{BERMAN-STEEL} who
stressed the importance of the non-conservation in the evolution of the
multipolar expansion terms of the atomic density matrix, for the observation
of sub-natural resonances in the NDFWM signal spectrum. Their work include a
detailed calculation, valid for degenerate atomic levels, of the nonlinear
atomic polarization to the lowest order in the incident fields. Recently, a
model allowing the numerical investigation of the complete response of a
degenerate two-level system (DTLS) to first order in the probe field for an
arbitrarily intense pump and arbitrary pump and probe polarizations \cite
{CSDTLS} was presented.

Recent experimental observations illustrate the need for a deeper
understanding of the role of the level degeneracy in the non linear response
of DTLS. An important example is provided by the spectroscopic response of
alkaline atoms in a magneto-optical trap (MOT). In this system the atoms are
submitted to the excitation of the trapping beams which are quasi resonant
with a closed degenerate transition. The absorption\cite{TABOSA,GRYNBERG}
and FWM\cite{HILICO,LOUNIS} spectra of magneto-optically trapped atoms
present characteristic features resulting from the level degeneracy.
Recently, the use of DTLS for efficient and highly selective NDFWM
generation in an atomic vapor was reported\cite{AKULSHINFWM}.

The aim of this paper is to address the role of the level degeneracy in
NDFWM on a closed transition between two degenerate atomic levels. The
dependence of NDFWM generation on the intensity of the pumping field and the
polarizations of the pump and probe will be analyzed. In the next section,
NDFWM spectra calculated for an homogeneous ensemble of DTLS under different
conditions are presented. The following section is devoted to the
experimental observation of NDFWM in cold cesium atoms. Discussion of the
experimental results in regard to the theoretical predictions follows.

\section{Theoretical predictions.}

\begin{figure}[tbp]
\begin{center}
\mbox{\epsfig{file=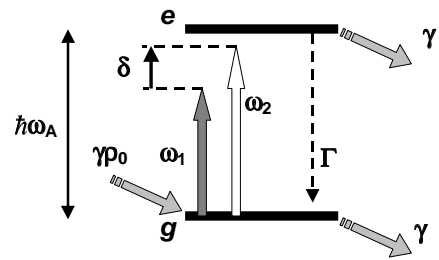,width=3.5in}}
\end{center}
\caption{Schematic representation of the configuration considered in the
theory. $e$ and $g$ are degenerate levels. Solid vertical arrows represent
the pump and probe fields (frequencies $\omega _1$ and $\omega _2$
respectively). Dashed arrow: decay by spontaneous emission. Oblique arrows:
arrival and departure of the atoms into and from the interaction zone. The
corresponding rates are indicated.}
\label{basic}
\end{figure}

The NDFWM generation in an homogeneous ensemble of DTLS is analyzed using
the semiclassical procedure already presented in \cite{CSDTLS}. In Fig. \ref
{basic} the essential elements of this model are reminded. The two-level
atoms are driven by a pump field of frequency $\omega _1$ of arbitrary
intensity and polarization. The {\em reduced} pump field Rabi frequency is $%
\Omega _1$. The driven atoms are probed by a weak field of frequency $\omega
_2$ $\equiv \omega _1+\delta $ and arbitrary polarization. Both fields are
nearly resonant with the atomic transition (transition frequency $\hbar
\omega _A$) between the ground level $g$ of total angular momentum $F_g$ and
the excited level $e$ of total angular momentum $F_e$. Having in mind the
experimental observation presented below, we will restrict our attention to
closed atomic transitions with $F_e=F_g+1$. Level $e$ decays back into $g$
through spontaneous emission at rate $\Gamma $. In addition to this
relaxation mechanism a state independent decay rate $\gamma $ ($\gamma \ll
\Gamma $) is assumed which represents the departure of the atoms from the
interaction zone. This departure is compensated at steady state by the
arrival (pumping term $\gamma \rho _0$) of ``fresh'' atoms isotropically
distributed in the ground level. ($\rho _0$ is the system density matrix in
the absence of any applied field). Since no other relaxation mechanism exist
for the ground state, $\gamma $ effectively plays the role of a ground state
relaxation rate. The calculation is exact to all orders in the pump field
and to first order in the probe field. The spectra represent the variation
of the squared modulus of the atomic polarization induced at the frequency $%
\omega _3\equiv 2\omega _1-\omega _2\equiv \omega _1-\delta $ as a function
of the probe to pump frequency offset $\delta $. Since the calculation
corresponds to the response of an homogeneous ensemble of atoms at rest, no
propagation effects such as phase matching, spatial emission pattern of the
generated wave or beam absorption (or amplification) are accounted for.

\begin{figure}[tbp]
\begin{center}
\mbox{\epsfig{file=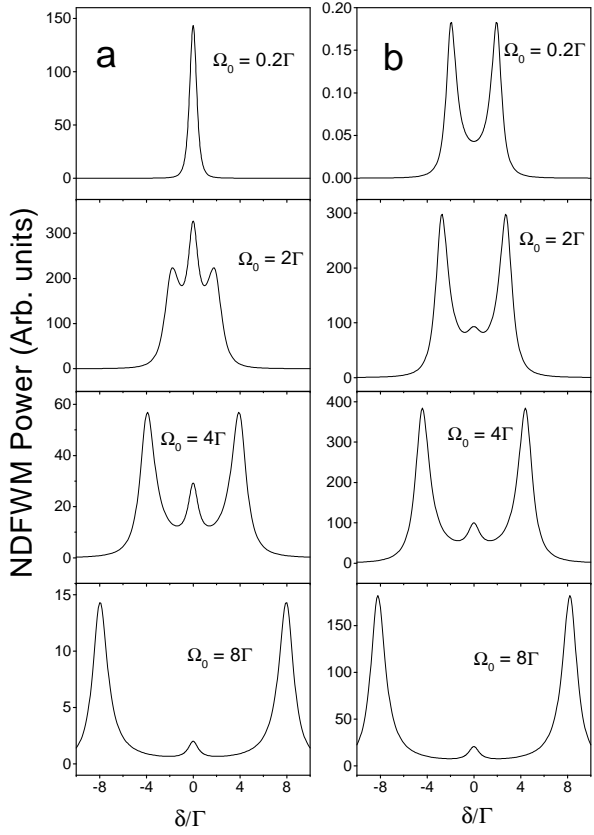,width=3.5in}}
\end{center}
\caption{Calculated NDFWM\ spectra for a pure two-level system. a) $\Delta
=0 $. b) $\Delta =2\Gamma $.}
\label{ptls}
\end{figure}

Before presenting the spectra calculated for DTLS, let us remind the main
features of the NDFWM spectra for PTLS\cite{BOYD,DUNCAN}. Figure \ref{ptls}
represent the spectra of the NDFWM power as a function of $\delta $ for $%
\Delta =0$ and $\Delta =2\Gamma $ ($\Delta \equiv \omega _A-\omega _1$) for
different values of the pump field Rabi frequency $\Omega _0$. We restrict
ourselves to closed transition where spontaneous emission is the only
relaxation mechanism. For all values of $\Omega _0$, the spectra present two
symmetric sidebands at $\delta =\pm \Omega $ where $\Omega \equiv \sqrt{%
\Omega _0^2+\Delta ^2}$ is the generalized Rabi frequency. The width of
these two peaks is determined by $\Gamma $. When $\Omega _0\gtrsim \Delta
,\Gamma $ a third peak develops at $\delta =0$ whose width is also
determined by $\Gamma $. Only when $\Omega \sim \Gamma $ the central peak
overcomes the two sidebands. As discussed in \cite{BOYD,DUNCAN}, the
relative amplitude of peak at $\delta =0$ is determined by the coherence
relaxation rate. In the present case this rate equals $\Gamma /2$ since only
radiative decay is assumed. When the transition is open and when the
escaping rates out of the two-level systems are different for the upper and
lower level, an additional peak appears at $\delta =0$ with a width
determined by the lower level relaxation rate\cite{DUNCAN}. Finally let us
mention that in PTLS the maximum NDFWM\ power is obtained for $\Omega _0\sim
\Gamma $.

\begin{figure}[tbp]
\begin{center}
\mbox{\epsfig{file=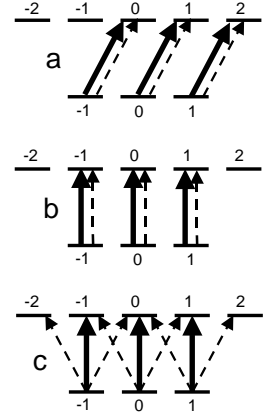,width=3.5in}}
\end{center}
\caption{Different level configurations corresponding to the transition $%
F_g=1\rightarrow F_e=2 $ for different choices of the pump and probe
polarizations. a) Circular and equal. b) Linear parallel. c) Linear
perpendicular. Solid (dashed) arrows correspond to the pump (probe) field.}
\label{levelscheme}
\end{figure}

We examine now the NDFWM power spectra for a DTLS with $F_g=1$ and $F_e=2$
(no magnetic field present). Three polarization cases were considered:
circular and equal pump and probe polarizations, parallel linear pump and
probe polarizations and linear and perpendicular pump and probe
polarizations. The corresponding level schemes are shown in Fig. \ref
{levelscheme} where an appropriate choice of the quantization axis is made
in each case.

\begin{figure}[tbp]
\begin{center}
\mbox{\epsfig{file=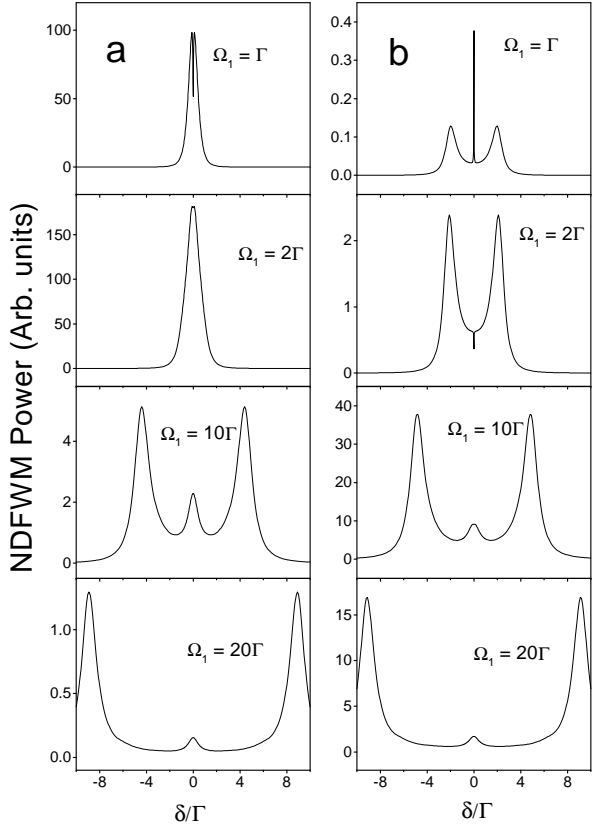,width=3.5in}}
\end{center}
\caption{Calculated NDFWM\ spectra for the transition $F_g=1\rightarrow
F_e=2 $ for circular and equal pump and probe polarizations ($\gamma
=0.01\Gamma $). a) $\Delta =0$. b) $\Delta =2\Gamma $.}
\label{ccd0d2}
\end{figure}

Figure \ref{ccd0d2} shows the spectra calculated for pump and probe fields
with the same circular polarization ($\sigma ^{+}$). As expected, due to the
optical pumping of the population towards the Zeeman sublevel in $g$ and $e$
with the highest magnetic quantum number (see Fig. \ref{levelscheme} a), the
system behaves as a PTLS in the limit of large $\Omega _1$. However,
striking differences appear for $\Omega _1\lesssim \Gamma $ in which case
the Zeeman optical pumping is only partial. Notice the presence at $\delta
=0 $ of narrow resonances: a peak for $\Omega _1=\Gamma $, $\Delta =2\Gamma $
and a dip for $\Omega _1=\Gamma $, $\Delta =0$ and $\Omega _1=2\Gamma $, $%
\Delta =2\Gamma $. The width of this narrow resonance is determined by $%
\gamma $. In this configuration, the pump and probe fields interact with
pairs of Zeeman sublevels of $e$ and $g$ with $m_e=m_g+1$ ($m_i$ is a
magnetic quantum number). Coupling between different pairs occurs only via
spontaneous emission. NDFWM signal arises from the coherent contribution of
different Zeeman sublevels pairs excited by the field. Due to spontaneous
emission, a given pair is not a close system in which case, a narrow
structure (width $\gamma $) is expected at $\delta =0$\cite{DUNCAN}. As
observed in Fig. \ref{ccd0d2} the sign of this narrow feature may vary
depending on the pump field intensity and detuning. It is the result of the
quantum interference of the contributions from the different Zeeman
sublevels pairs to the non linear atomic polarization. A similar situation
occurs in the case of parallel linear polarizations discussed next.

\begin{figure}[tbp]
\begin{center}
\mbox{\epsfig{file=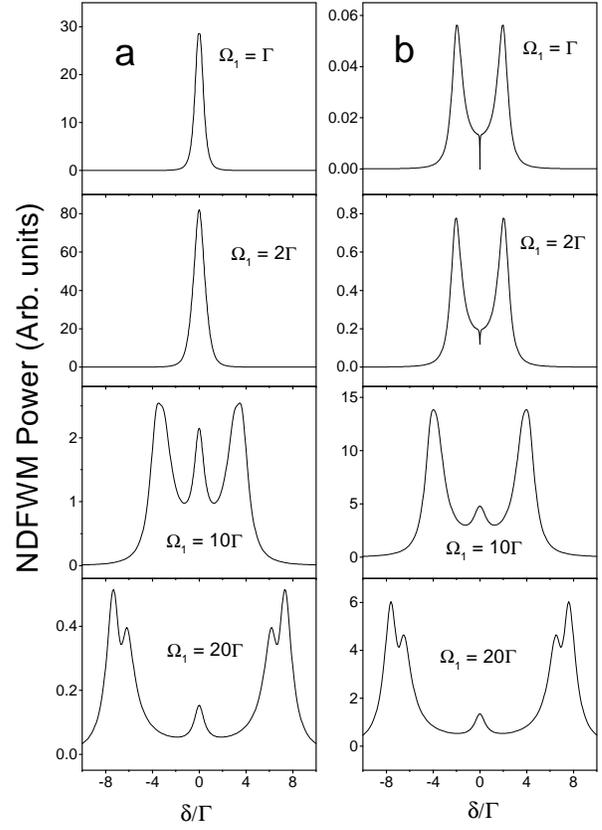,width=3.5in}}
\end{center}
\caption{Calculated NDFWM\ spectra for the transition $F_g=1\rightarrow
F_e=2 $ for linear and parallel pump and probe polarizations ($\gamma
=0.01\Gamma $). a) $\Delta =0$. b) $\Delta =2\Gamma $.}
\label{lld0d2}
\end{figure}

\begin{figure}[tbp]
\begin{center}
\mbox{\epsfig{file=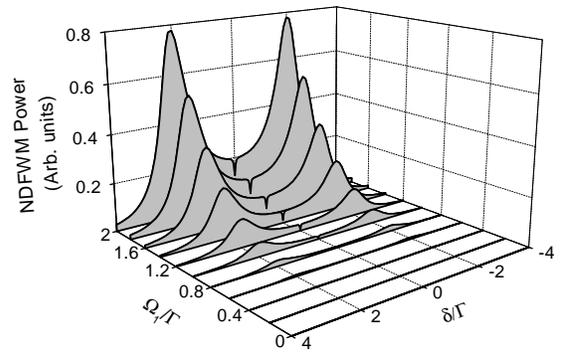,width=3.5in}}
\end{center}
\caption{NDFWM power spectra for the transition $F_g=1\rightarrow F_e=2$
with $\Delta =2\Gamma $ and $\gamma =0.01\Gamma $ for different values of
the pump field Rabi frequency $\Omega _1$.}
\label{allms}
\end{figure}

\begin{figure}[tbp]
\begin{center}
\mbox{\epsfig{file=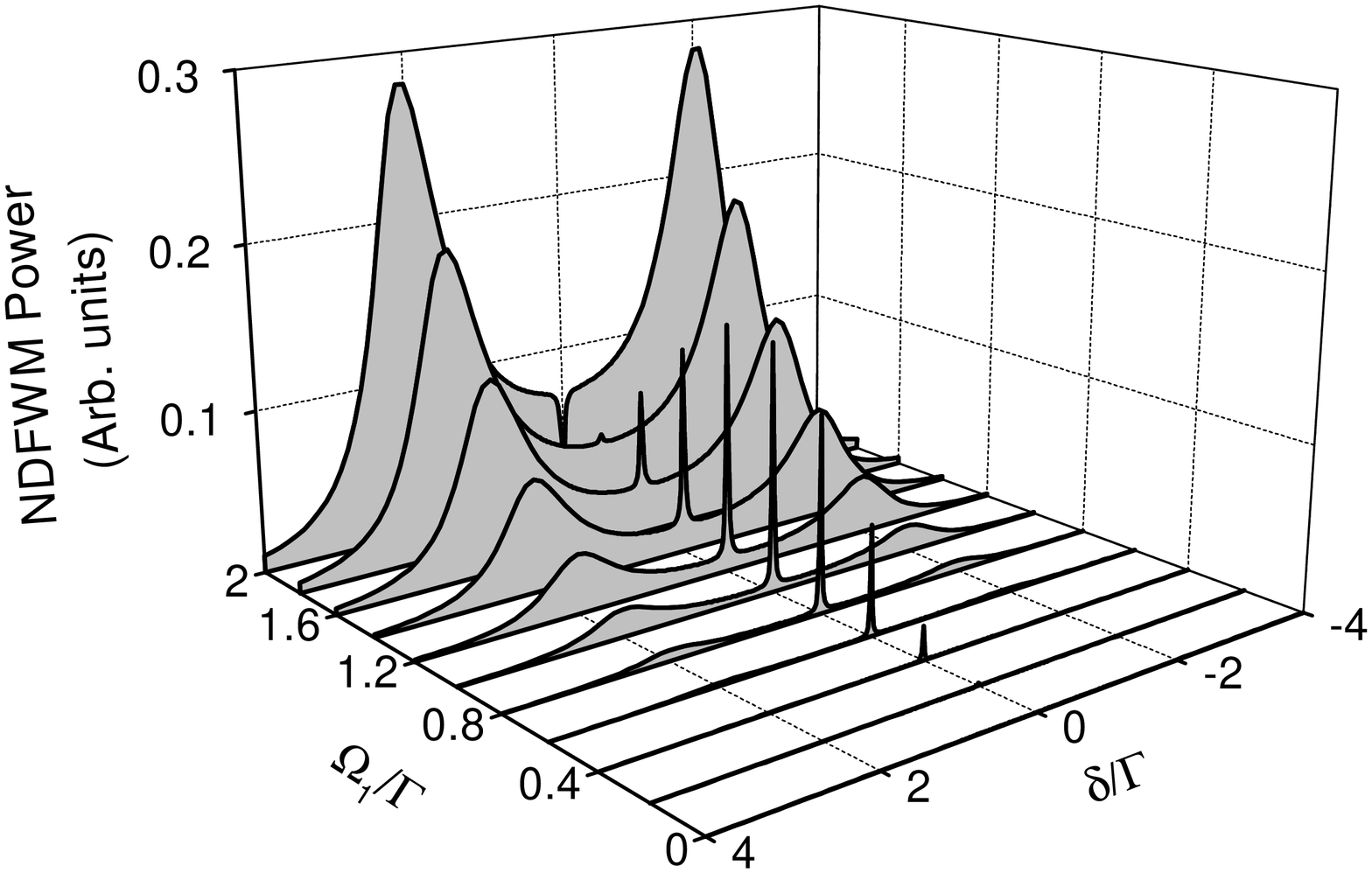,width=3.5in}}
\end{center}
\caption{Square modulus of the contribution to the atomic polarization at
frequency $\omega _3\equiv \omega _1-\delta $ from the pair of Zeeman
sublevels with $m=0$.}
\label{mo}
\end{figure}

\begin{figure}[tbp]
\begin{center}
\mbox{\epsfig{file=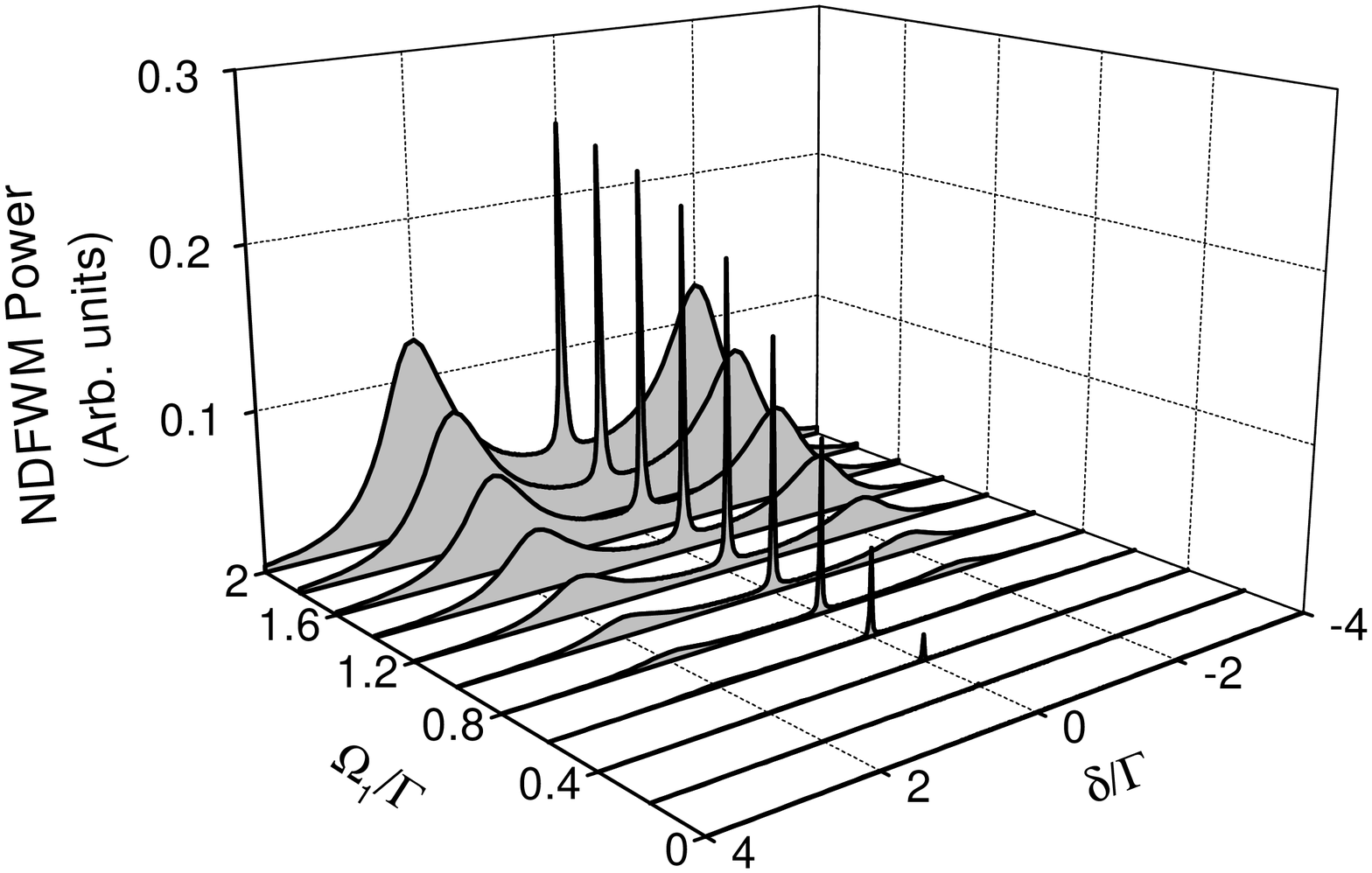,width=3.5in}}
\end{center}
\caption{Square modulus of the contribution to the atomic polarization at
frequency $\omega _3\equiv \omega _1-\delta $ from the pairs of Zeeman
sublevels with $\left| m\right| =1$.}
\label{m1}
\end{figure}

Figure \ref{lld0d2} refers to the spectra calculated for pump and probe
fields with parallel linear polarizations. Unlike the previous case, in the
limit of large $\Omega _1$, the system does not behave as a PTLS. Instead,
taking as the quantization axis the direction of the common pump and probe
polarization, it should be seen as three ($2F_g+1$) two-level systems linked
through spontaneous emission (Fig. \ref{levelscheme} b). Each two-level
system consist of a two Zeeman sublevels with $m_g=m_e$. Since the strength
of the atom field coupling dependents only on $\left| m_g\right| $, two
different values of the pump Rabi frequency occur in the present example.
This explains the splitting observed in the two sidebands for large $\Omega
_1$. At low pump intensities a narrow resonance (width determined by $\gamma 
$) appears at $\delta =0$ for $\Delta =2\Gamma $. Also a narrow dip is
present at $\delta =0$ for $\Delta =0$ and $\Omega _1<\Gamma $ (not shown in
the figure). The appearance of narrow features for zero pump to probe
detuning with a width determined by the effective ground state relaxation
rate was predicted for open two-level systems\cite{DUNCAN}. In the present
case this feature should be seen as the consequence of the population
transfer, due to spontaneous emission, between pairs of levels connected by
the fields. A detailed look into the evolution of this narrow resonance with
the pump field power (for $\Delta =2\Gamma $) is presented in Fig. \ref
{allms}. The resonance at $\delta =0$ is negative (dip) for $\Omega
_1>0.4\Gamma $. This result is to be compared with the contribution to the
nonlinear atomic polarization arising from the different pairs of Zeeman
sublevels with $m_e=m_g$. These contributions can be extracted from the
calculation after identification of the density matrix coefficients
corresponding to a given value of $\left| m_i\right| $ ($i=e,g$). Figs. \ref
{mo} and \ref{m1} represent the square modulus of the nonlinear atomic
polarization contributions arising from the Zeeman sublevels pairs with $%
\left| m_g\right| =0$ and $\left| m_g\right| =1$ respectively. Only in the
case of $\left| m_g\right| =0$ and for the largest value of $\Omega _1$ the
central resonance is opposite to the lateral sidebands. In all other cases
this resonance has the same sign than the lateral sidebands. This is always
the case for $\left| m_g\right| =1$. Consequently, the dip observed at $%
\delta =0$ in the NDFWM spectra in Fig. \ref{allms} is the result of
(destructive) quantum interference between the two contributions coming from 
$\left| m_g\right| =0,1$. Indeed, the two contributions have opposite phase
over all the considered range of $\Omega _1$.

\begin{figure}[tbp]
\begin{center}
\mbox{\epsfig{file=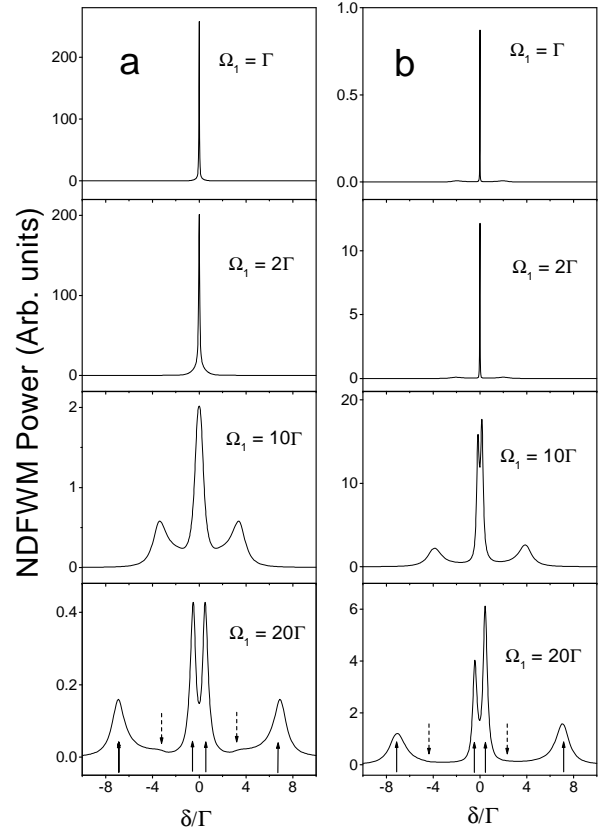,width=3.5in}}
\end{center}
\caption{Calculated NDFWM\ spectra for the transition $F_g=1\rightarrow
F_e=2 $ for perpendicular linear pump and probe polarizations ($\gamma
=0.01\Gamma $). a) $\Delta =0$. b) $\Delta =2\Gamma $.}
\label{lpld0d2}
\end{figure}

\begin{figure}[tbp]
\begin{center}
\mbox{\epsfig{file=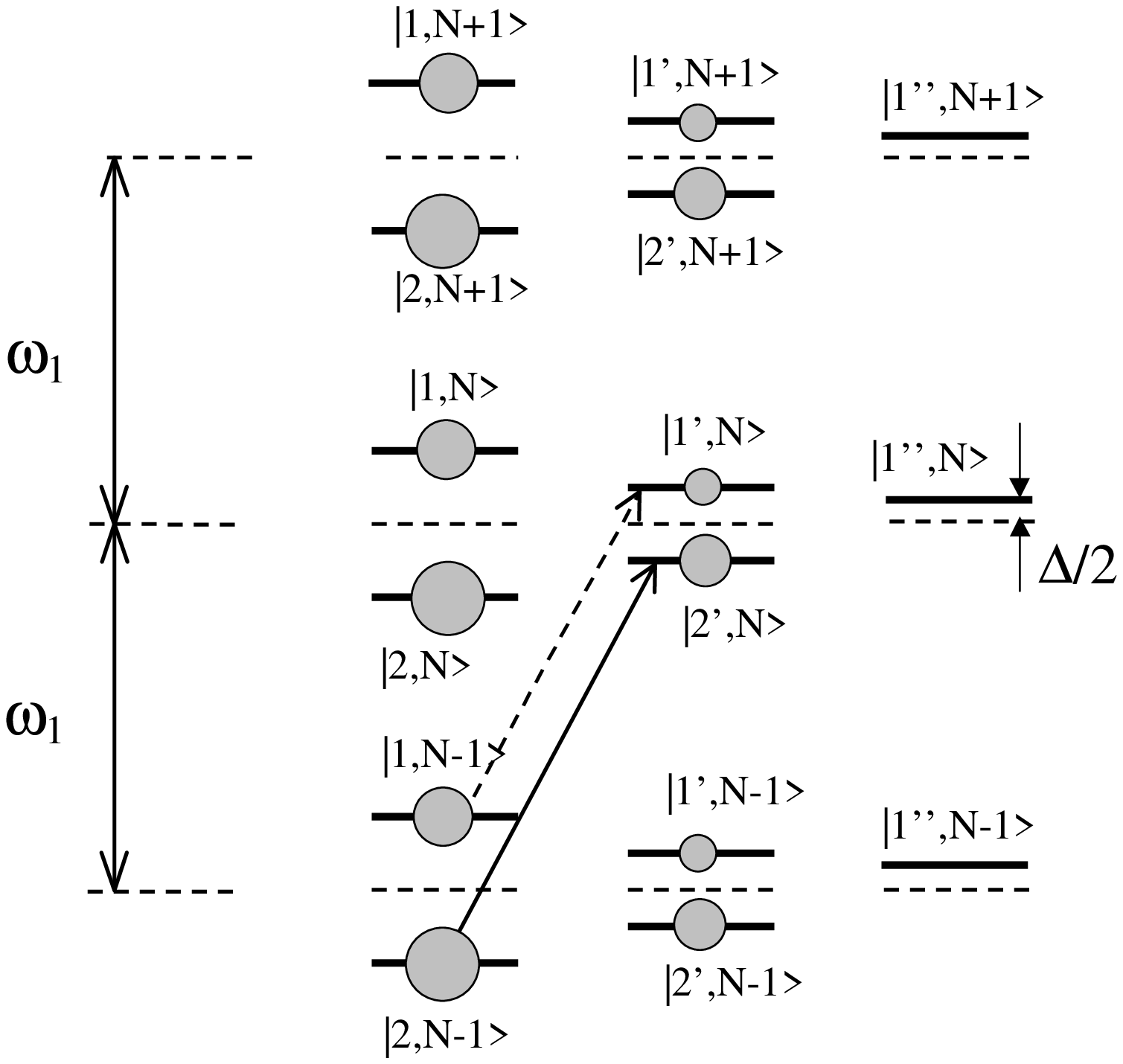,width=3.5in}}
\end{center}
\caption{Dressed atom energy level scheme for a transition $F_g=1\rightarrow
F_e=2$ driven by a $\pi $ polarized pump field. The notations are taken from
Ref. \protect\cite{CSDTLS}. The arrows illustrate two possible probe field
transitions giving rise to resonances in the NDFWM spectra corresponding to
opposite values of $\delta$.}
\label{dressed}
\end{figure}

The NDFWM spectra obtained for perpendicular linear pump and probe
polarizations (Fig. \ref{lpld0d2}) are rather different to the ones
presented above\cite{LAM}. For all the considered pump field intensity range
the spectra are dominated by the features occurring around $\delta =0$. In
fact for $\Omega _1\lesssim 2\Gamma $ the spectral sidebands are barely
visible and the spectra is mainly composed by a large central resonance with
a width of the order of $\gamma $ \cite{CSDTLS}. This configuration produces
for $\Omega _1\approx \Gamma $ the largest NDFWM yield\cite{AKULSHINFWM}.
For increasing $\Omega _1$ the central peak splits into two components. Also
in this case the spectra are asymmetric for $\Delta \neq 0$. The main
features of these spectra can be qualitatively understood using the dressed
states picture of the degenerate atomic system in the presence of the pump
field. A similar analysis was carried in \cite{MOLLOW}(Appendix) to examine
the probe absorption spectra of driven DTLS in this configuration. The
dressed-state energy level scheme for a $F_g=1\rightarrow F_e=2$ transition
driven by a $\pi $ polarized pump is presented in Fig. \ref{dressed}
following the conventions adopted in \cite{MOLLOW}. NDFWM resonances are
expected to occur when the probe field is resonant with a transitions
between dressed levels that are coupled to the pump field photons. Four
different probe frequencies satisfy this condition. The corresponding values
of $\delta $ (see Table II in \cite{MOLLOW}) are indicated with solid arrows
in the two lower spectra of Fig. \ref{lpld0d2}. They correspond to the main
features in the NDFWM\ spectra. From a simple dressed-state analysis, no
resonance is expected to occur for the probe frequencies corresponding to
transitions ending in the $\left| 1",N\right\rangle $ dressed-atom levels
since these states are not coupled to the pump photons due to the $m_e=m_g$
selection rule valid for a $\pi $ polarized pump. Nevertheless small
features appear in the calculated spectra at these positions (dashed arrows
in Fig. \ref{lpld0d2}). They are due to non-secular terms usually neglected
in the dressed atom approach.

\section{Experimental observation of NDFWM in cold cesium.}

\begin{figure}[tbp]
\begin{center}
\mbox{\epsfig{file=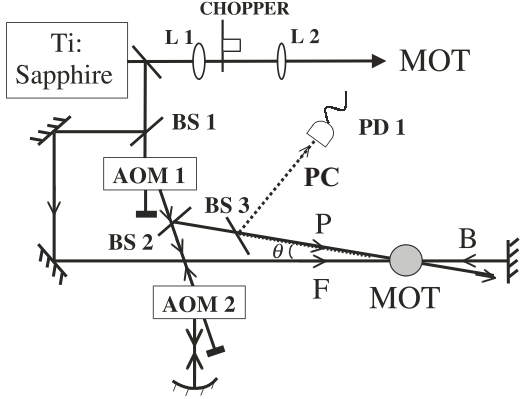,width=3.5in}}
\end{center}
\caption{$a)$ Experimental setup. AOM: acousto-optic modulator. BS: beam
splitter. PD: photodiode. MOT: magneto-optically trapped atomic sample.}
\label{setup}
\end{figure}

The observation of the NDFWM spectra was performed in a sample of cold
cesium atoms produced in a magneto-optical trap (MOT). Light from a
Ti:sapphire laser, nearly resonant with the cesium cycling transition $%
6S_{1/2}(F=4)\rightarrow 6P_{3/2}(F=5)$, was employed both for the trapping
of the atoms and for the investigation of the NDFWM. The experimental setup
is shown schematically in Fig. \ref{setup}. The frequency of the Ti:sapphire
laser is red-detuned by approximately two natural linewidth $(\Gamma /2\pi
=5.3\ MHz)$. A repumping diode laser, not shown in the figure, recycles the
population lost to the $6S_{1/2}(F=3)$ ground state. We use a backwards FWM
configuration, where the two counter-propagating pump fields, the forward
(F) and backward (B), have the same frequency and the same linear
polarization. The probe beam (P), is linearly polarized and makes a small
angle ($\theta =4^o$) with the pumping beams. The probe beam has its
frequency scanned around the frequency of the pump beams with the help of
two acousto-optic modulators as shown in Fig. \ref{setup}. The relative
polarization between the pump and the probe fields is controlled by a
half-wave plate. The generated (nearly) phase conjugated signal (PC), which
propagates in opposite direction with respect to beam P, is reflected out of
a $50/50$ beam splitter and detected by a fast photodiode. The trapping
beams are switched on and off by a mechanical chopper with a transmission
duty cycle of $95\%$. The number of cold atoms was estimated by measuring
the absorption of the probe beam and is of order of $10^7$. The NDFWM
spectra were recorded within a $\sim 1\ ms$ time interval during which the
trapping beams were blocked and the quadrupole magnetic field was turned
off. Each of the F and B pump beams have an intensity of $7\ mW/cm^2$. The
probe beam intensity is approximately equal to $0.7\ mW/cm^2$. No
significant modification of the spectra was observed at lower probe beam
intensities. The maximum NDFWM\ power generated was of the order of $1\ \mu
W $. Typical spectra, recorded as a function of the pump to probe frequency
offset $\delta $, are shown in Figs. \ref{expteoll}a) and \ref{expteolpl}a)
for the probe beam polarization respectively parallel and perpendicular to
the pump beam polarization.

\section{Discussion.}

The comparison of the observed spectra with the theoretical predictions
presented above is not direct. While the calculation applies to a
homogeneous ensemble of atoms at rest, the PC signal is the result of
cooperative emission of atoms under different excitation conditions. In the
experiment the pump field, at a given position in the sample, is due to the
combined incidence of the F and B beams. The counterpropagating geometry of
these beams produces a standing wave, consequently, the pump field intensity
is spatially modulated. In general the pump field polarization may be
spatial dependent. However, in the present study we have restricted
ourselves to the case where the beams F and B have the same linear
polarization and consequently no spatial variation of the polarization
occurs. Also, the PC signal is sensitive to propagation effects in the
atomic sample which are responsible for the phase matching condition and for
possible spatial dependent amplification or depletion of the NDFWM through
the medium. Finally, one should remind that the phase matching condition
imposes constraints on the polarization components that may be present in
the PC beam (only transverse components). However, in the two polarization
cases considered in the experiment, linear parallel and linear perpendicular
pump and probe polarizations, due to symmetry, the nonlinear atomic
polarization should be parallel to the probe polarization (in the absence of
magnetic field). In consequence, all polarization components of the NDFWM\
field are able to propagate along the PC beam.

In order to allow the comparison between the theoretical prediction and the
experimental observation, we have incorporated into the calculation the
intensity distribution of the pump field in the atomic sample. Propagation
effects were not considered. Such approach, expected to be valid for an
optically thin sample, is only approximative in our case since the peak
probe absorption is around $50\%$. Assuming that the F and B beams produce a
perfect standing wave, we have considered a sine wave distribution of $%
\Omega _1$ in the interval $0\leq \Omega _1\leq \Omega _{1MAX}$. For each
value of $\Omega _1$, the NDFWM field was calculated assuming that a steady
state is reached. The total NDFWM field was taken as the sum of the
contributions for each $\Omega _1$. Finally the NDFWM power correspond to
the square modulus of the total field. The calculations were performed for
an $F_g=4\rightarrow F_e=5$ transition with $\Delta =-2\Gamma $
(corresponding to the experimental conditions) and $\gamma =0.01\Gamma $. No
magnetic field was considered. The value of $\Omega _{1MAX}$ was adjusted to
fit the experimental spectra. The best agreement was obtained for $\Omega
_{1MAX}=18\Gamma $.

\begin{figure}[tbp]
\begin{center}
\mbox{\epsfig{file=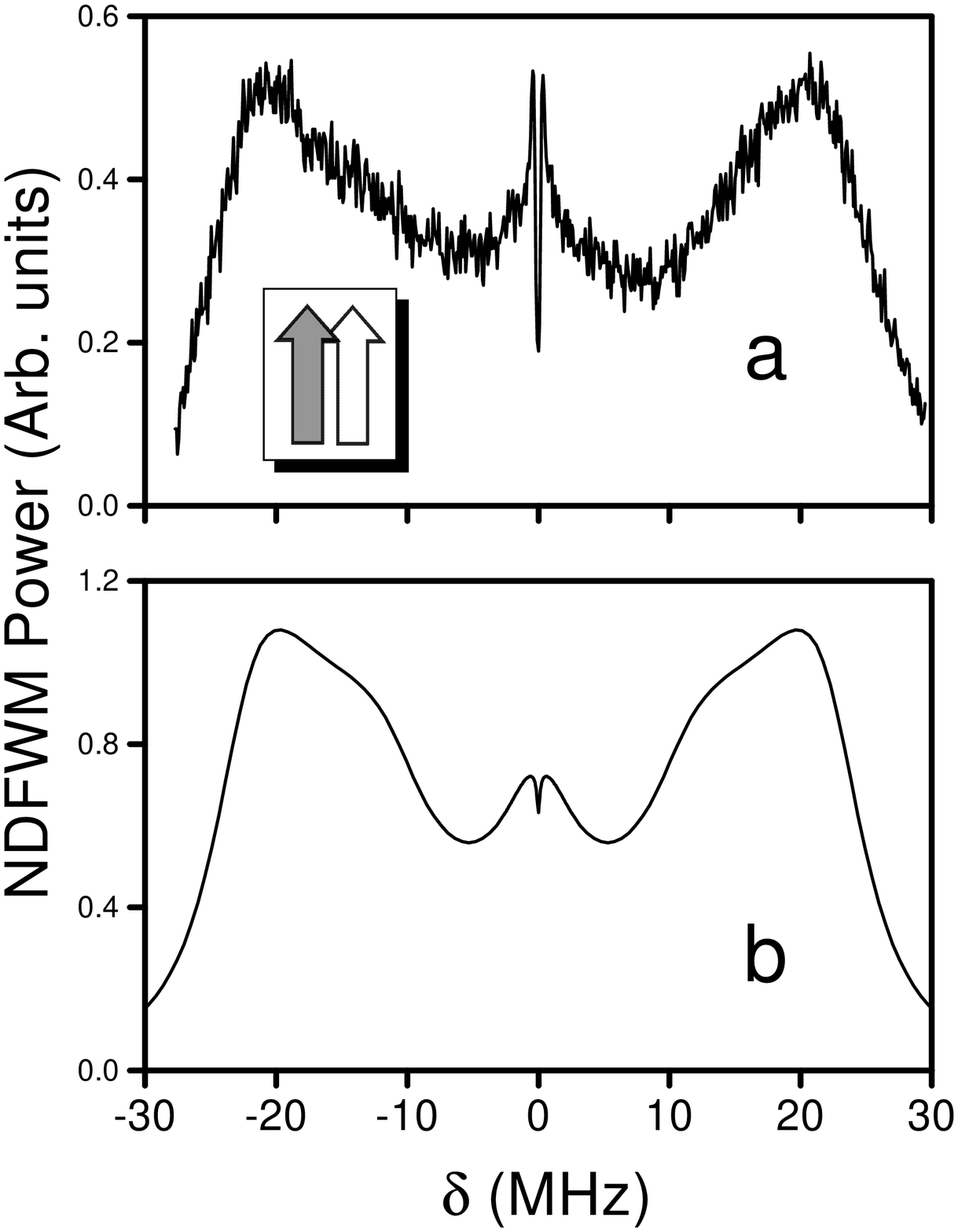,width=3.5in}}
\end{center}
\caption{Observed $a)$ and calculated $b)$ NDFWM spectra for linear and
parallel pump and probe polarizations. The vertical axes scales are
independent (the same scales are used in Fig.\ref{expteolpl}).}
\label{expteoll}
\end{figure}

Figure \ref{expteoll}b) represents the calculated spectrum for linear and
parallel pump and probe polarizations with the parameters corresponding to
the experimental conditions. The spectrum is dominated by the two sidebands.
The width of the sidebands is mainly due to the inhomogeneity of the Rabi
frequency. The largest values of $\Omega _1$ are responsible for the central
peak presenting a width of the order of $\Gamma $ while the atoms
corresponding to $\Omega _1\lesssim \Gamma $ are responsible for the narrow
dip present at $\delta =0$. The main features of the experimental spectrum
are well reproduced. Some differences appear in the amplitude and shape of
the narrow resonance around $\delta =0$. According to the theoretical
considerations presented above, this narrow resonance is the consequence of
the coupling between different Zeeman sublevels pairs through spontaneous
emission. Its width is governed by the time of flight relaxation rate $%
\gamma $ which effectively plays the role of a ground-state decay rate. For
cold atoms, the average time of flight across the exciting beams is rather
long (more than $1\ ms$ for a $1\ mm$ diameter beam) and the corresponding
value of $\gamma <3\times 10^{-5}\ \Gamma $ too small to account for the
observed resonance. However, for typical MOT temperatures, the average time
os flight across one spatial period of the stationary wave produced by the F
and B beams is three orders of magnitude shorter and correspond to $\gamma
\sim 10^{-2}\ \Gamma $ as assumed in the calculation. At this point one
should notice that a short travelling time across the standing wave pattern
of the pump field in not compatible with our assumption of a steady state
reached for each value of $\Omega _1$. This suggests that a more
sophisticated theoretical approach, incorporating the atomic motion, would
be more appropriate for a precise description of the spectra. In addition to
the finite interaction time, another mechanism that can affect the narrow
resonance around $\delta =0$ is the possible leakeage out of the closed
two-level transition. In our case, this may occur though non resonant
excitation of other excited state hyperfine levels. However the
corresponding rate can be estimated to be smaller than $10^{-3}\Gamma $.
Also, calculations carried for open transitions give spectral shapes very
different from those observed in Fig. \ref{expteoll}.

\begin{figure}[tbp]
\begin{center}
\mbox{\epsfig{file=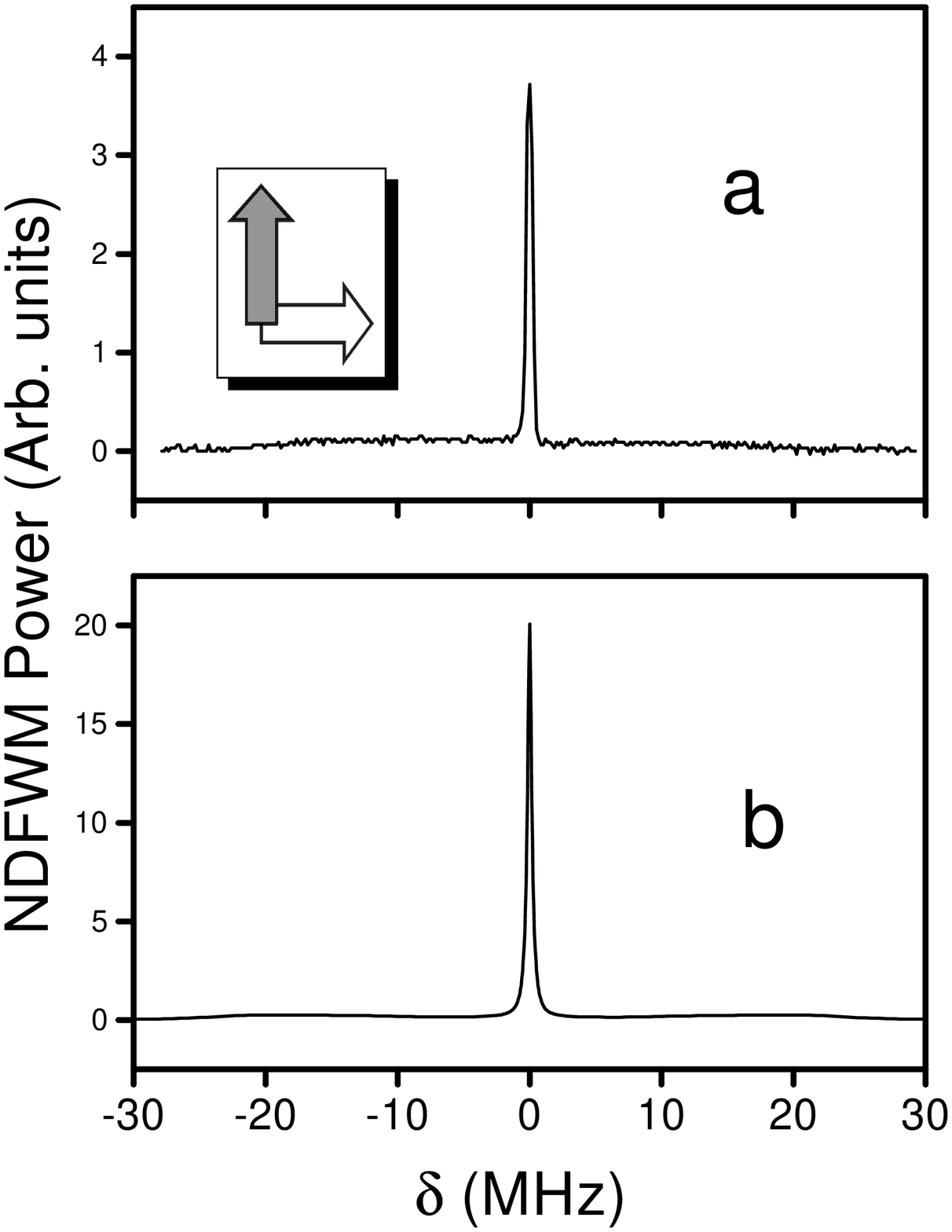,width=3.5in}}
\end{center}
\caption{Observed $a)$ and calculated $b)$ NDFWM spectra for linear and
perpendicular pump and probe polarizations. The vertical axes scales are
independent (the same scales are used in Fig.\ref{expteoll}).}
\label{expteolpl}
\end{figure}

The comparison between observed and calculated spectra in the case of linear
and perpendicular pump and probe polarizations are presented in Fig. \ref
{expteolpl}. The main features of the spectra are well reproduced by the
calculation. Notice the significant increase in the maximum NDFWM yield with
respect to the previous case. The spectrum is in the present case dominated
by the narrow resonance at $\delta =0$. The width of this resonance is also
given at low pump intensities by $\gamma $. As already pointed in \cite
{CSDTLS} and \cite{AKULSHINFWM}, this configuration provides the largest
NDFWM yield. The difference in the spectral profile existing between Figs. 
\ref{expteoll} and \ref{expteolpl} constitutes a clear demonstration of the
essential role of the Zeeman degeneracy and optical polarizations in the
NDFWM process.

\section{Conclusions.}

The process of NDFWM in a closed atomic transitions with $F_e=F_g+1$ has
been examined both theoretically and experimentally. Large differences in
the spectra are observed for different choices of the exciting fields
polarizations revealing the crucial role of the internal level structure on
the non-linear process. The spectra present distinct features which are
determined by the different relaxation rates and characteristic frequencies
of the system ($\Gamma ,\ \gamma ,\ \Omega _1,\ \Delta $)\cite{BERMAN-STEEL}%
. NDFWM was observed in a cold sample of cesium atoms in a PC experiment for
two different choices of the pump and probe polarizations. In spite of the
difference existing between the experimental conditions and the assumptions
of the theory a good agreement between calculated and observed spectra was
obtained. Nevertheless, some features of the spectra, associated to the
longest relaxation processes, indicate the need for a more detailed
theoretical approach including the atomic motion and spatial field
distribution.

\section{Acknowledgments.}

The authors acknowledge fruitful discussions with J.R. Rios Leite. This work
was supported by CNPq (PRONEX), CAPES and FINEP (Brazilian agencies) and by
CSIC, CONYCIT and PEDECIBA (Uruguayan agencies).

\begin{references}
\bibitem{NICO}  N. Bloembergen, {\it Nonlinear Optics}, Benjamin, New York
(1965).

\bibitem{SHEN}  Y.R. Shen, The principles of nonlinear optics, J. Wiley \&
Sons, New York (1984).

\bibitem{ROTHBERG}  L.J. Rothberg and N. Bloembergen, Phys. Rev. A {\bf 30},
820 (1984). L.J. Rothberg and N. Bloembergen, Phys. Rev. A {\bf 30}, 2327
(1984).

\bibitem{BOYD}  R.W. Boyd, M.G. Raymer, P. Narum and D.J. Harter, Phys. Rev.
A {\bf 24}, 411 (1981).

\bibitem{DUNCAN}  D.G. Steel and J.T. Remillard, Phys. Rev. A {\bf 36}, 4330
(1987).

\bibitem{BERMAN-STEEL}  P.R. Berman, D.G. Steel, G. Khitrova and J. Liu,
Phys. Rev. A {\bf 38}, 252 (1988).

\bibitem{ANDERSEN}  O.K. Andersen, D. Lenstra and S. Stolte, Phys. Rev. A 
{\bf 60}, 1672 (1999).

\bibitem{CSDTLS}  A. Lezama, S. Barreiro,. A. Lipsich and A.M. Akulshin,
Phys. Rev. A {\bf 61}, 013801 (2000).

\bibitem{TABOSA}  J.W.R. Tabosa, G. Chen, Z. Hu, R.B. Lee and H.J. Kimble,
Phys. Rev. Lett. {\bf 66}, 3245 (1991).

\bibitem{GRYNBERG}  D. Grison, B. Lounis, C. Salomon, J.Y. Courtois and G.
Grynberg, Europhys. Lett. {\bf 15}, 149 (1991).

\bibitem{HILICO}  L. Hilico, P. Verkerk and G. Grynberg, C. R. Acad. Sci.
Paris {\bf 315}, 285 (1992).

\bibitem{LOUNIS}  B. Lounis, P. Verkerk, C. Salomon, J.Y. Courtois and G.
Grynberg, Nonlinear Optics {\bf 5}, 459 (1993).

\bibitem{AKULSHINFWM}  A.M. Akulshin, S.Barreiro and A. Lezama, Quantum
Electronics {\bf 30}, 189 (2000); [Kvantovaya Elektronika {\bf 30}, 189
(2000)].

\bibitem{LAM}  J.F. Lam, D.G. Steel and R.A. McFarlane, Phys. Rev. Lett. 
{\bf 56}, 1679 (1986).

\bibitem{MOLLOW}  A. Lipsich, S. Barreiro, A.M. Akulshin and A. Lezama,
Phys. Rev. A {\bf 61}, 053803 (2000).
\end{references}
\end{document}